\magnification 1200
\hsize=31pc             
\vsize=55 truepc         
\baselineskip=26 truept 
\hfuzz=2pt               
\vfuzz=4pt               
\pretolerance=5000        
\tolerance=5000        
\parskip=0pt plus 1pt 
\parindent=16pt

\def\var{\partial}

\vskip 1.truecm\noindent
\centerline {\bf EXACT SOLUTION OF A 2D RANDOM ISING MODEL}
\vskip 1.4truecm
\noindent
\centerline{Maurizio Serva}
\vskip .4truecm
\centerline{\it
 Dipartimento di Matematica,  Universit\`a dell'Aquila and  I.N.F.M.}
\centerline{\it I-67010 Coppito, L'Aquila, Italy}
\vskip .7truecm
\centerline{ABSTRACT}
\vskip .4truecm
The model considered is a $d=2$ layered random Ising system
on a square lattice with nearest neighbours interaction. 
It is assumed that all the vertical couplings are equal and
take the positive value $J$  
while the horizontal couplings are quenched random variables
which are equal in the same row but can take the
two possible values $J$ and $J-K$ in different rows.
The exact solution is obtained in the limit case $K\to\infty$
for any distribution of the horizontal couplings.
The model which corresponds to this limit
can be seen as an ordinary Ising system 
where the spins of some rows, chosen at random,
are frozen in an antiferromagnetic order.
No phase transition is found
if the horizontal couplings are independent
random variables while for correlated disorder one finds
a low temperature phase with
some glassy properties.

\noindent
\medskip
PACS NUMBERS: 05.50.+q, 02.50.-r 
\vfill\eject

\bigskip
\bigskip

Ising spin glasses have been solved exactly 
in their mean field version [1,2] while
as far as I know no exact solutions are 
available at finite dimensionality $d \geq 2$.
Indeed, in presence of disorder
even a $d=1$ system with magnetic field is 
a very complicated problem [3]
and compact exact solutions can be found only 
in special cases [4].
This is frustrating since
it is not always clear if the qualitative results
of a mean field approximation 
are shared by the finite dimension model. 
In this paper I am far from answering 
this problem, nevertheless
I exactly solve a class of $d=2$ layered random Ising systems 
which in some conditions have a low temperature phase.
The nature of this non-ferromagnetic
low temperature phase is still unclear to me,
but there are some indications that it shares some of 
the properties of a glassy phase.

The models I consider are defined as follows:
the interaction is effective only
between nearest neighbours on a square lattice;
all the vertical couplings are equal
while the horizontal couplings are quenched variables which
are equal in the same row but can take different values in  
different rows;
the vertical couplings take the positive value $J$  
while the horizontal couplings $J_i$ can take the
two possible values $J$ and $J-K$ with $K \to \infty$.
These models have frustration since the product of
the signs of the coupling around a plaquette can be negative.
Different distributions of the horizontal couplings
correspond to different models of the class; in the simplest case the 
$J_i$ are independent random variables and take the value 
$J$ with probability $1-p$ and $J-K$
with probability $p$.

Layered Ising models of this type have been first considered
by B.M. McCoy and T.T. Wu [5,6] for the non frustrated case
which is used for studying the effect of
quenched randomness on the ferro-para transition.
These authors deal with the determinant which occurs in the 
Pfaffian approach and while they do not provide an explicit 
exact solution of the problem  
they are able to show that the free energy has an infinitely 
differentiable singularity at the transition.
Layered models with frustration have been 
studied by R. Shankar and Ganpathy Murthy [7],
not only their topic but also
their approach is the same of this work since they deal with the 
row to row transfer matrices. 
They do not find out an exact solution, nevertheless
they map the problem into a collection 
of $d=1$ random field Ising systems 
from which they can extract a lot of informations.
In particular they provide evidence for the
existence of a low temperature phase.
 
Let me now state more precisely the problem.
Assume that $N=LM$ is the number of spins,
$L$ is the number of rows and $M$ the number of columns,
the hamiltonian can be written as
$$
H_N = -\sum_{ij}
\left( J \sigma_{i,j}\sigma_{i+1,j}
+J_i \sigma_{i,j}\sigma_{i,j+1}
\right)
\eqno(1)
$$
where the $J_i$ are the horizontal couplings
whose value only depend on the row $i$ 
and not on the column $j$.
One can write $ J_i=J-\eta_i K$
where the quenched variables $\eta_i$
can take the value $0$ and $1$
according to a given distribution.
In the independent case 
$\eta_i =0$ with probability $1-p$ and $\eta_i=1$
with probability $p$.
The partition function is
$$
Z_N = \sum_{\{\sigma\}} \exp
\{ \sum_{ij}\beta (J\sigma_{i,j}\sigma_{i+1,j}
+J\sigma_{i,j}\sigma_{i,j+1}
-\eta_i K (1+\sigma_{i,j}\sigma_{i,j+1}))\}
\eqno(2)
$$
where the constant term $\sum_{ij} \eta_i K $
has been added to the hamiltonian in order to avoid
divergences in the $K\to\infty$ limit.
After having defined $\Gamma \equiv J\beta$ and performed the limit
$K \to \infty$ one obtains
$$
Z_N = \sum_{\{\sigma\}} \prod_{ij}
\left[
\exp\{ \Gamma \sigma_{i,j}\sigma_{i+1,j}
+\Gamma \sigma_{i,j}\sigma_{i,j+1}\}
\left(1-{1+\sigma_{i,j}\sigma_{i,j+1} \over 2}{\eta_i}\right)
\right]
\eqno(3)
$$
The terms in parenthesis equal $1$ when $\eta_i =0$ and
$(1-\sigma_{i,j}\sigma_{i,j+1})/2$ when $\eta_i =1$.
Notice that in this second case the antiferromagnetic order
between neighbour spins on the row is imposed,
in fact, if $\sigma_{i,j}$ and $\sigma_{i,j+1}$
have the same sign they give a vanishing contribution
to the partition function.
It is now clear that (3) defines 
a class of Ising model with both
vertical and horizontal couplings equal to $J$ and 
with the spins of some rows frozen in an antiferromagnetic order.
The frustration comes out from the fact that 
the tendency to the ferromagnetic
alinement due to the positive
couplings is in competition with the tendency
to the antiferromagnetic alinement induced by the 
frozen spins on the unfrozen ones.
A similar problem, where the spin are randomly frozen
in a random direction has been solved in $d=1$ in [4],
and studied in $d=2$ at zero temperature in [8].

The advantage of considering layered disorder is
that one can apply a standard diagonalization method 
[7,9], and reduce the problem to the evaluation 
of the trace of products of random matrices.
Following the same steps of [7,9] one easily finds
the free energy
$$
f=-{J \over 2\Gamma}\log (2\sinh 2\Gamma)
- {J\over 2\pi\Gamma }\int_0^\pi \gamma(q,\Gamma)dq
\eqno(4)
$$

where  

$$
\gamma(q,\Gamma)=
\lim_{L\to\infty} {1\over L}\log {\rm Tr}\prod_{i=1}^L 
T_i(q,\Gamma)
\eqno(5)
$$ 
The $2 \times 2$ matrices $T_i(q,\Gamma)$ can be written as the product 
$T_i(q,\Gamma)=E_i \,\cdot  T(q,\Gamma)$ where the 
$$
E_i=\left( \matrix{1&0\cr 0&1-\eta_i\cr}\right)
\eqno(6)
$$ 
are random and equal the identity when $\eta_i =0$ and the
up projector $\tau^+ = (1+\tau_3)/2$ when $\eta_i=1$.
The matrix $T(q,\Gamma)$, on the contrary, is constant
and reads
$$
T(q,\Gamma)=\exp\{-\Gamma\tau_3\}
\exp\{2\Gamma^*(\tau_3\cos q +\tau_1\sin q )\}
\exp\{-\Gamma\tau_3\}
\eqno(7)
$$ 
where $\tau_1$ and $\tau_3$ are Pauli matrices and
$\Gamma^* \equiv -{1\over2}\log( \tanh \Gamma)$.

The trace of a product of random matrices
is easily accessible via computer simulation
but it cannot be, in general, exactly computed.
In the present case, nevertheless, following
a similar method as in [4], it is possible to find out the
compact analytical result.
Consider a given realization
of the quenched variables $\eta_i$ 
and look at the product of matrices in (5).
Since $T_i(q,\Gamma)=E_i \,\cdot  T(q,\Gamma)$ 
that product reduces to a product of 
matrices $T(q,\Gamma)$ and up projectors $\tau^+$.
The first and the second $\tau^+$ will be separated
by $l_1$ matrices $T(q,\Gamma)$,
the second and the third by $l_2$ matrices $T(q,\Gamma)$, and so on.
The $l_n$ are random variable which can take
the values $1,2,....$  whose distribution 
can be easily found out once  
the distribution of the $\eta_i$ is given.
The order number $n$ goes from $1$ to 
$n_f=L/\bar{l}$,
in fact, one must have $\sum_{n=1}^{n_f} l_n =L$
so that $\sum_{n=1}^{n_f} l_n /n_f \equiv \bar{l}=L/n_f$.
With the help of these considerations 
one can rewrite (5) as 
$$
\gamma(q,\Gamma)=
\lim_{L\to\infty} {1\over L} \log\prod_{n=1}^{L/\bar{l}} 
[T(q,\Gamma)^{l_n}]_{11}
=\lim_{L\to\infty} {1\over L} \sum_{n=1}^{L/\bar{l}} 
\log[T(q,\Gamma)^{l_n}]_{11}
\eqno(8)
$$
where $[T(q,\Gamma)^{l_n}]_{11}$ is the up left entry
of $T(q,\Gamma)^{l_n}$.
If  $P(l)$ is the
probability that two successive rows of infinitely negative
couplings are separated by $l$ rows of finite positive couplings
than $\bar{l}\equiv \sum_{l=1}^\infty \, l \, P(l)$ and
$$
\gamma(q,\Gamma)=
\sum_{l=1}^\infty {1\over \bar{l}}P(l)\log[T(q,\Gamma)^{l}]_{11}
\eqno(9)
$$
In order to find the explicit form for (9) it is convenient
to write
$$
T(q,\Gamma)=\exp\{\epsilon(\tau_3\cos\phi+
\tau_1\sin\phi)\}
\eqno(10)
$$
where
$$
\cosh\epsilon=
{\cosh^2 2\Gamma \over \sinh 2 \Gamma}-\cos q
\eqno(11)
$$
and
$$
\cos\phi=
{\cosh 2 \Gamma (\cos q-\sinh 2 \Gamma)
\over (\sin^2q+\cosh^2 2 \Gamma ( \cos q - \sinh 2 \Gamma)^2)^{1 \over 2}}
\eqno(12)
$$
Using (10) it is immediate to obtain 
$$
[T(q,\Gamma)^{l}]_{11}=\cosh(l\epsilon)+ \cos\phi\sinh(l\epsilon)
\eqno(13)
$$
Finally:
$$
f=-{J \over 2\Gamma}\log (2\sinh2\Gamma)
- {J\over 2\pi\Gamma\bar{l} }\sum_{l=1}^\infty P(l)
\int_0^\pi \log(\cosh(l\epsilon)+ \cos\phi\sinh(l\epsilon))dq
\eqno(14)
$$
where $\epsilon$ and $\phi$ are given in (11) and (12).

The probability $P(l)$ for the simplest choice
of independent $\eta_i$ is 
$P(l)=p(1-p)^{l-1}$ and $\bar{l}=1/p$.
In this case one can prove that 
the system has no phase transition,
except for $p=0$ where it trivially reduces
to the ordinary Ising model.
In Fig. 1 it is shown the specific heat $C$
in correspondence of different values of $p$;
one can notice that the logarithmic divergence is smoothed
showing the absence of transition.
Nevertheless, the model is frustrated and its
zero temperature properties are
not completely trivial. 
One can compute the $T=0$
energy $f_0$ end entropy $s_0$
and finds
$$
f_0=-2J(1-p)^2
\eqno(15)   
$$
$$
s_0=Jp^2 (1-p) \log \left( {\sqrt{5}+1 \over 2} \right)
\eqno(16)
$$
$s_0$ is not vanishing for $p\neq0,1$
showing an exponential degeneration of the ground state
due to the frustration of the model.

Since the transition disappears for 
$p\neq0$ 
the role of $p$ reminds that of
a magnetic field which also 
suppresses the transition.
The analogue of the spontaneous magnetization is
obtained in the limit $p\to0$ as
$$
f'\equiv
\left[{\var f \over \var p}\right]_{p=0}=
- {J\over 2\pi\Gamma }
\int_0^\pi \log\left( {1+ \cos\phi \over 2} \right)dq
\eqno(17)
$$
This quantity is continuous while its derivative
${d f' \over d T}$ is not as shown in Fig. 2
where one can see a logarithmic divergence at $T_c$
(the Onsager critical temperature).

The circumstance that a phase transition can be found only at 
$p=0$ suggests to look more carefully at the model
around this value.
If one chooses $p = \alpha/L$ one has a vanishing 
$p$ in the thermodynamic limit
and the free energy is the same of that of the standard Ising model.
Nevertheless, one has a random finite number of
frozen rows. This number is Poisson distributed with
intensity $\alpha$ and it is different for different
realizations of the disorder (no self-averaging).
The distance between two given frozen rows is also a 
random number of order of $L$ and it also varies
from a realization to another.
The final result is that the frozen rows separate
a random number of regions of random size of order $N$
whose magnetization at $T\leq T_c$ is $\pm m(T)$ 
independently one from the other ($m(T)$ 
is the Onsager spontaneous magnetization at temperature $T$).
As a consequence of this fact,
the whole system can be in all the states corresponding
to all the possible combinations of magnetization of each region.
In conclusion,
one has the same free energy 
of a standard Ising system 
but a number of pure states each of
them corresponding to a different 
local magnetization.
The situation is completely analogous to that studied in
[10] for a diluted $d=1$ model at zero
temperature. Following the same line of [10]
it is easy to compute the overlap probability,
in particular for large $\alpha$ one has
$$
P(q) \simeq
{1 \over \sqrt{2\pi t}}\exp\{- {q^2 \over 2t} \}    
\eqno(18)
$$
where $t=m(T)^2/\alpha$.
(18) implies that the overlap between two different pure states vanishes
in the limit $\alpha \to \infty$ ,
nevertheless it should be noticed that
the self-overlap $q_{max} = m(T)^2$,
being independent on $\alpha$, remains finite.

To summarize: for $p\neq0$
there is no phase transition, while for $p=\alpha/L$
one has a glassy like phase
which comes out from an artificial construction
which maintain the same free energy of the Ising model.
It is straightforward, at this point, to
look at an intermediate situation, where the number of frozen rows
is of order of $L$ but they can be much more separated than in the 
independent case.
This task can be accomplished with the choice
$P(l) =a/l^3$ ($a$ is the normalization constant)
which replaces the exponential distribution
$P(l)=p(1-p)^{l-1}$ of the independent case.
By substituting this expression in (14) 
one can easily compute the free energy and look
at the eventual divergences. 
In spite of the fact that
the free energy
is now different from that of the standard Ising model 
one still finds a phase transition at the Onsager temperature.
Nevertheless this phase transition does not correspond to a divergence
in the specific heat,
but in its derivative ${d C \over d T}$
which is plotted in Fig. 3.
I have not been able to 
quantitatively characterize the low phase temperature
with an order parameter.
In this phase, in fact, while the spontaneous magnetization 
vanishes,
both the overlap and a parameter connected with
the  antiferromagnetic
order seems to differ from zero.
I hope that some light on this point  will 
come out from future research.

I acknowledge the financial support of the I.N.F.N., 
National Laboratories of Gran Sasso ({\it Iniziativa Specifica} FI11)
and the hospitality of the Institut Superieure Polytechnique 
de Madagascar.
I thank Andrea Crisanti, Umberto Marini Bettolo Marconi and 
Julien Raboanary for useful discussions and suggestions.

\vfill\eject
\noindent
{\bf References}
\bigskip
\bigskip
\item{ [1]}
D. Sherrington and S. Kirkpatrick,
 Phys. Rev. Lett. {\bf 32}, (1975), 1792
\bigskip
\item{ [2]}
M. Mezard, G. Parisi and M. Virasoro, {\it Spin glass theory and beyond},
 World Scientific Singapore 1988
\bigskip
\item{[3]}
G. Paladin and M. Serva,
Phys. Rev. Lett,  {\bf 69}, (1992), 706 
\bigskip
\item{[4]}
G. Grinstein and D. Mukamel,
Phys. Rev. B {\bf 27}, (1983), 4503
\bigskip
\item{[5]}
B.M. McCoy and T.T. Wu, 
Phys. Rev. {\bf 76}, (1968), 631
\bigskip
\item{[6]}
B.M. McCoy, 
Phys. Rev. B. {\bf 2}, (1970), 2795
\bigskip
\item{[7]}
R. Shankar and Ganpathy Murthy,
Phys. Rev. B  {\bf 36}, (1987), 536
\bigskip
\item{[8]}
B. Derrida and Y. Shnidman,
J. Phisique Lett. {\bf 45}, (1984), L-507
\bigskip
\item{[9]}
T. D. Schultz, D. C. Mattis and E. H. Lieb
Rev. Mod. Phys. {\bf 36},(1964), 856
\item{[10]}
A. Crisanti, G. Paladin, M. Serva and A. Vulpiani, 
Journal de Physique 1 {\bf 3}, (1993), 1993
\bigskip

\vfill\eject
\centerline {\bf Figure Captions}
\vskip 0.5truecm
\noindent
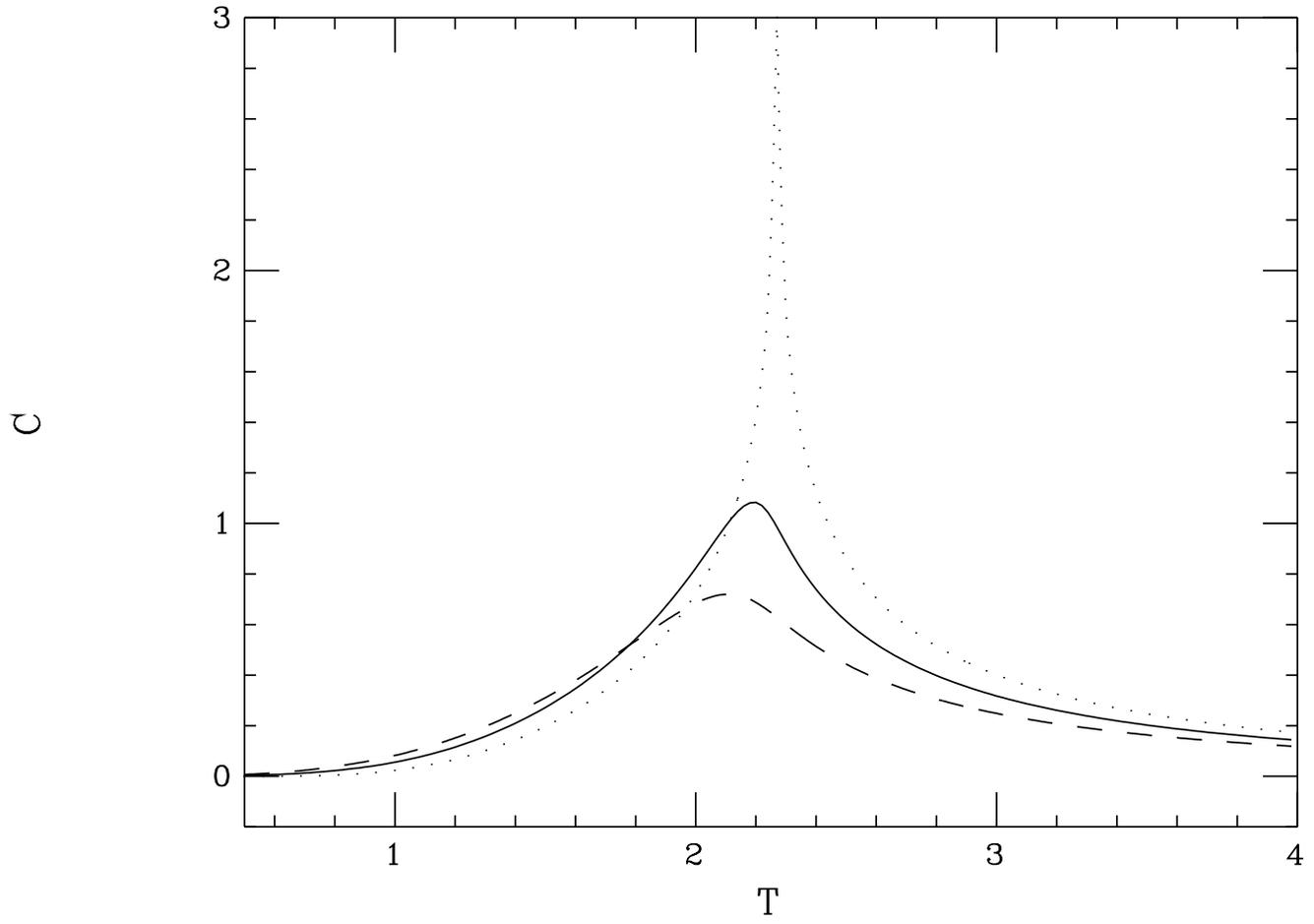
\item{Fig. 1}
Specific heat $C$ as function of the temperature $T$.
The dotted line corresponds to the Ising model ($p=0$),
the full line to $p=0.1$ and
the dashed line to $p=0.2$
\bigskip
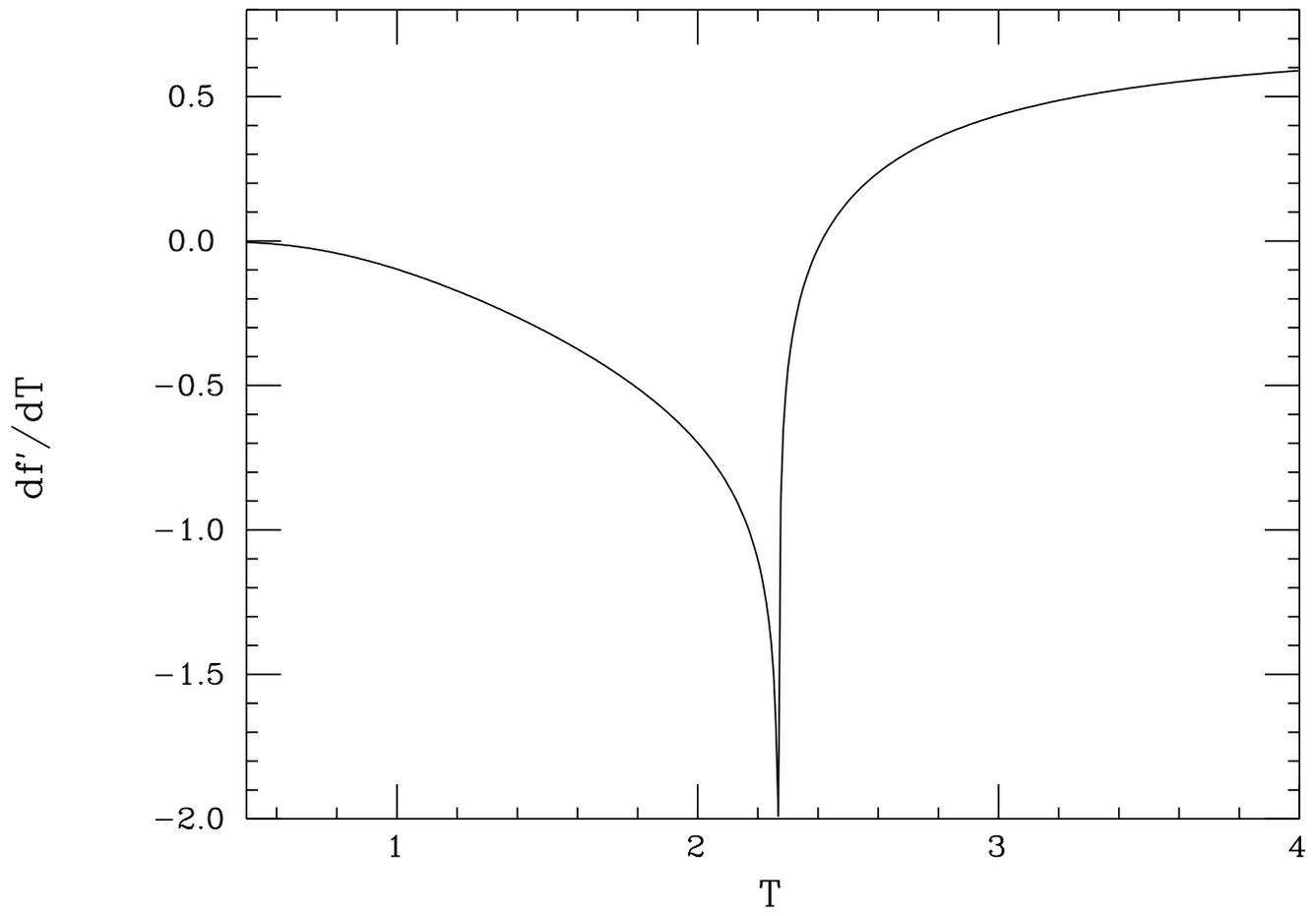
\item{Fig. 2}
Temperature derivative ${df'\over dT}$
of $f'\equiv
\left[{\var f \over \var p}\right]_{p=0}$ 
as function of $T$
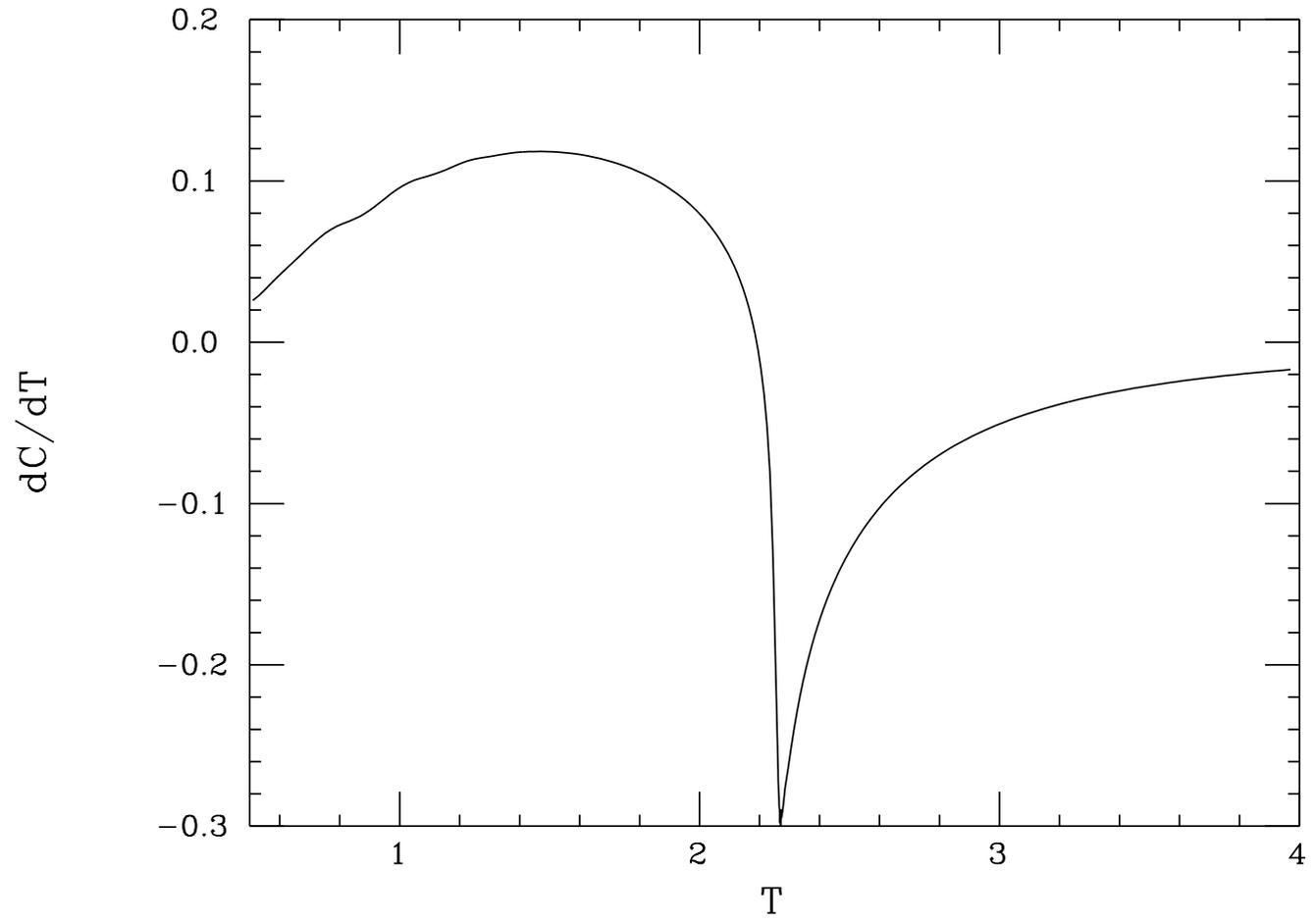
\item{Fig. 3}
Temperature derivative ${dC\over dT}$
of the specific heat
for the $P(l)=a/l^3$ model.
\bye